# First Keck Nulling Observations of a Young Stellar Object: Probing the Circumstellar Environment of the Herbig Ae star MWC 325

S. Ragland[1,†], K. Ohnaka[2], L. Hillenbrand[3], S. T. Ridgway[4], M. M. Colavita[5], R. L. Akeson[6], W. Cotton[7], W. C. Danchi[8], M. Hrynevych[9], R. Millan-Gabet[6], W. A. Traub[5]

[1]W. M. Keck Observatory, [2]Max-Planck-Institut für Radioastronomie, [3]California Institute of Technology, [4]National Optical Astronomy Observatories, [5]Jet Propulsion Laboratory, California Institute of Technology, [6]NASA Exoplanet Science Institute, [7]National Radio Astronomy Observatory, [8]NASA Goddard Space Flight Center, [9]MRO, New Mexico Tech

**ABSTRACT**

We present the first $N$-band nulling plus $K$- and $L$-band $V^2$ observations of a young stellar object, MWC325, taken with the 85 m baseline Keck Interferometer. The Keck nuller was designed for the study of faint dust signatures associated with debris disks, but it also has a unique capability for studying the temperature and density distribution of denser disks found around young stellar objects. Interferometric observations of MWC 325 at $K$, $L$ and $N$ encompass a factor of five in spectral range and thus, especially when spectrally dispersed within each band, enable characterization of the structure of the inner disk regions where planets form. Fitting our observations with geometric models such as a uniform disk or a Gaussian disk show that the apparent size increases monotonically with wavelength in the 2–12 μm wavelength region, confirming the widely held assumption based on radiative transfer models, now with spatially resolved measurements over broad wavelength range, that disks are extended with a temperature gradient. The effective size is a factor of about 1.3 and 2 larger in the $L$-band and $N$-band, respectively, compared to that in the $K$-band. The existing interferometric measurements and the spectral energy distribution can be reproduced by a flat disk or a weakly-shadowed nearly flat-disk model, with only slight flaring in the outer regions of the disk, consisting of representative "sub-micron" (0.1 μm) and "micron" (2 μm) grains of a 50:50 ratio of silicate and graphite. This is marked contrast with the disks previously found in other Herbig Ae/Be stars suggesting a wide variety in the disk properties among Herbig Ae/Be stars.

## 1. INTRODUCTION

The study of disks around pre-main sequence (PMS) stars supports a long-term goal of the detection and characterization of exoplanets. The initial formation of a circumstellar disk is a consequence of the angular momentum distribution during star formation. For the first several million years, a PMS star is surrounded by a disk of gas and dust, which is a remnant reservoir left over from the build-up of stellar mass during the early stage of star formation. As the central star ages the disk disappears as a consequence of: (1) accretion of disk material on to the central star, (2) ejection of material through stellar winds, (3) irradiation by stellar and other energetic photons and (4) formation of planets. As circumstellar disks provide the raw material for planet formation, clues to the physical conditions of planet formation can be inferred from a detailed characterization, with the terrestrial planet zone corresponding to the inner regions of young stellar object (YSO) disks. Improved observations of inner disk evolution helps us to form a better model and hence improve our understanding of planet formation.

---

[†] Email: sragland@keck.hawaii.edu

Herbig Ae/Be (HAeBe) stars are a class of PMS stars of intermediate mass. The spectral types of HAeBe range from B to F, and their masses range from 2–8 $M_\odot$ (Herbig 1960). The lifetime of the primordial disks of HAeBe stars is less than ~3 Myr (Hillenbrand et al. 1992; Hernandez et al. 2005). These stars may be the progenitors of Vega-type debris stars.

A large number of HAeBe stars, across the luminosity range, have now been spatially resolved at near-infrared wavelengths using long baseline interferometry (Millan-Gabet et al. 1999, 2001; Danchi et al. 2001; Eisner et al. 2003, 2004, 2007, 2009, 2010; Leinert et al. 2004; Monnier et al. 2005, 2006; Malbet et al. 2007; Tatulli et al. 2007; Acke et al. 2008; Isella et al. 2008; Kraus et al. 2008a,b; Tannirkulam et al. 2008a,b; Ragland et al. 2009; Benisty et al. 2010a,b).

We have been carrying out a systematic investigation of YSO disks of different luminosity types through multi-color interferometry using the Keck Interferometer (KI). With ~3 mas angular resolution at 2.2 µm, the Keck Interferometer can resolve the inner circumstellar regions (< 1 AU for distances <300 pc) of PMS star disks where terrestrial planets can form. Our first results on MWC 419, based on simultaneous *K*- & *L*-band measurements, were published in Ragland et al. 2009, where we found that MWC 419 exhibits relatively flat disk geometry. Prior to our work, multicolor interferometry results in the *H*- or *K*- and *N*-bands were reported for only three other YSO disks, MWC 297 (Acke et al. 2008), MWC 147 (Kraus et al. 2008a) and HD 100546 (Benisty et al. 2010b) from observations with the Very Large Telescope Interferometer (VLTI). In this paper we report observations of MWC 325 in *K*-band (2.0–2.4 µm), *L*-band (3.5–4.1 µm) and *N*-band (8–12 µm).

MWC 325 (V1295 Aql; HD 190073; HIP 98719) is classified as A2 IVev (Mora et al. 2001). An alternate spectral type of A2IIIpe can also be found in the literature (Schütz et al. 2009). The distance to MWC 325 is uncertain. The most commonly adopted distance in the literature has been 290 pc, based on the lower limit from Hipparcos measurements (Van der Ancker et al. 1998). However revised Hipparcos measurements (van Leeuwen 2007) suggest a lower limit of 340 pc. More recently, Montesinos et al. (2009) estimate a distance of $767^{-76}_{+139}$ pc along with other stellar parameters such as mass, age and luminosity through an iterative SED fitting process in conjunction with PMS tracks and isochrones. Their reported values for mass, luminosity, and age are $5.05^{-0.47}_{+0.54} M_\odot$, $470.8^{-88.2}_{+189.2} L_\odot$, and $0.6^{-0.2}_{+0.2}$ Myr, respectively. These authors adopt an effective temperature and *log g* of 9500K and 3.37 ± 0.08, respectively. Catala et al. (2007) adopt an effective temperature, *log g*, and *log(L/L$_\odot$)* of 9250 ± 250K, 3.5 ± 0.5, and 1.9 ± 0.12, respectively, and obtain stellar mass, stellar radius, and age of 2.85 ± 0.25 $M_\odot$, 3.6 ± 0.5 $R_\odot$, and 1.2 ± 0.6 Myr, respectively, using theoretical evolutionary tracks. In this paper, we adopt an effective temperature, luminosity, radius, mass and distance of 9250 K, 83 $L_\odot$, 3.6 $R_\odot$, 2.85 $M_\odot$, and 340 pc, respectively, for the central star. The reason for adopting a distance of 340 pc is that our Kurucz model fits to BVRI photometric data from literature favors this value over the estimate of Montesinos et al. (2009) for a star on the main sequence. We discuss the impact of this choice on our results in Section 3.2.

MWC 325 shows a low projected stellar rotational velocity, reported to be 9 km/s (Acke & Waelkens 2004) to 12 km/s (Pogodin et al. 2005). The expected rotational velocity for an A2 IV/III star is in the range 100-150 km/s (Tassoul 2000). The observed low velocity suggests that the central star has either a low intrinsic rotational velocity or a typical intrinsic velocity but is viewed close to face-on.

MWC 325 exhibits a classical P Cygni II-type profile for the Balmer lines, H$_\alpha$ to H$_\delta$ (Pogodin et al. 2005). These authors conclude that MWC 325 exhibits a strong stellar wind with an optically thick equatorial disk. PAH emission was not detected in MWC 325 from a 3 μm survey of HAeBe stars (Acke & van den Ancker 2006).

In this paper we report spectrally resolved *K-, L-* and *N*-band interferometric measurements of MWC 325. These probe physically different regions of the circumstellar disk having representative temperatures ~1400K, ~800K, and ~300K, respectively for simple blackbody grains. Moreover, these multi-wavelength observations probe different spatial scales as the fringe spacing on the sky is directly proportional to the wavelength. Thus, the *N*-band observations probe spatial scales roughly 5 times larger than that of the *K*-band observations. Discrete spatial distributions such as dust-rims and relatively-smooth spatial distributions such as classical accretion disks are expected to have different size-verses-wavelength behaviors, and can be distinguished in such multi-color observations at well separated wavelengths. MWC 325 has been spatially resolved previously by interferometers in the *H*- & *K*-bands (Millan-Gabet et al. 2001; Eisner et al. 2004, 2007, 2009; Monnier et al. 2006). These previous measurements enable a check for time variability in inner disk radii, which might be expected given the dynamical time scale in this active region of the disk.

In Section 2, we present our observations and data reduction. In Section 3, we describe the data analysis where we fit the visibility-squared ($V^2$) data with various YSO disk models. In Section 4, we discuss our results in conjunction with previous interferometric observations and in Section 5, we provide a brief summary.

## 2. OBSERVATIONS AND DATA REDUCTION

The KI is a near- and mid-infrared long-baseline interferometer consisting of two 10 m diameter telescopes separated by an 85 m baseline at a position angle of ~38° east of north, with the ability to null a central point source using phase control of the interfering beams. Both Keck telescopes are equipped with adaptive optics systems designed to compensate for atmospheric-induced wavefront aberrations, a crucial mitigation element for large-aperture long-baseline optical interferometry. The spatial resolution of our KI observations ($\lambda/2B$) is ~ 2.7 mas, ~4.5 mas and ~12 mas in the *K-*, *L-*, and *N*-band respectively. The spectral resolution is R ~ 24, 56, and 21 over the *K-*, *L-*, and *N*-bands, respectively.

The observable in the *K,* and *L*-bands is visibility-squared $V^2$ (the squared modulus of the visibility containing information about the spatial extent of the source) as a function of wavelength. The $V^2$ data in the *K-* and *L*-bands are collected simultaneously in the dual-band mode described by Ragland et al. 2008, 2009, and we refer to these as K/L data, below. In this mode, the Keck telescope pupils are split into halves, each with a separate beam train (this is an adaptation of the nuller configuration, described next) over the 85 m baseline. One beam train feeds a *K*-band $V^2$ system, which was the first KI science instrument, while the other feeds a more recently commissioned *L*-band $V^2$ system. Both systems operate similarly, with the observational scenario incorporating interleaved scans of calibrators, in addition to per-scan calibrations of the background and the flux ratio between the two apertures. The only significant difference between them is that, for *L*-band, because of the higher thermal background, nods to dark sky are required for the background measurement in order to achieve an accurate photometric calibration.

The *N*-band data are collected with the Keck Interferometer Nuller (KIN) described by Colavita et al. 2008, 2009. The KIN operates differently than an ordinary $V^2$ instrument, with a different observable: the null leakage *l*. However, for relatively compact objects of angular extent small compared to the short baseline fringes (400 mas), the leakage is simply transformed to $V^2$ using the following equation (Koresko et al. 2006; Colavita et al. 2009) for comparison with the K/L data.

$$V^2_{measured} = \left[\frac{1-l}{1+l}\right]^2 \tag{1}$$

$$\delta V^2_{measured} = 4\left[\frac{(1-l)}{(1+l)^3}\right]\delta l$$

Equation 1 assumes that the entire disk is much smaller than the KIN cross fringe, which has fringe spacing of 400 mas, and is a good approximation for these data.

The KIN uses the split pupil mode with two nulling beam combiners on the long 85 m baselines. In the data collection configuration, these two combiners are stabilized on a destructive fringe, ideally canceling all on-axis (point source) light, and only transmitting the extended emission. The leakage from these two nulling beam combiners is combined in a third beam combiner, which uses fast fringe scanning to measure the leakage flux in the presence of the large *N*-band background. By changing the track point on the nulling beam combiners from destructive interference to constructive interference, a "photometric" measurement can made to normalize the leakage flux, producing the main observable, the (normalized) null leakage *l*. Because the achievable control bandwidth using just the *N*-band light is lower than needed to compensate for atmospheric turbulence, the KIN also uses two *K*-band systems on the long baselines for fringe stabilization. However, as these systems provide *K*-band $V^2$ (with a resolution of 4 channels across the *K*-band) as an auxiliary data product, both *K*- and *N*-band data can be collected simultaneously, and we refer to these as K/N data, below. The observational scenario for the KIN is similar to other modes, using interleaved calibrators to measure the system leakage, in the same way interleaved calibrators measure the system visibility for $V^2$ observations.

The field-of-view of the *K*- and *L*-band instruments is defined by the single mode fibers that couple light to the HAWAII (*K*-band) and PICNIC (*L*-band) infrared detectors, and that of the KIN is defined by the pinhole used in the *N*-band camera. The resultant field-of-view (FWHM) is ~60 mas for *K*-band, ~100 mas for the *L*-band and ~ 500 mas for *N*-band measurements. These field restrictions were used in our modeling work.

The observations reported here were taken on the nights of UT 19 August 2008 (*K*- and *N*-band measurements) and UT 27 October 2009 (*K*- and *L*-band measurements). We observed five calibrators – HD 188310, HD 193579, HD 206445, HD 190007 and HD187691 - under similar observing conditions as the science target. We performed bracketed calibration - meaning that our observing sequence consisted of calibrator-target-calibrator measurements. The adopted angular diameters of the calibrators are 2.0 ± 0.2, 2.0 ± 0.2, 1.8 ± 0.2, 0.5 ± 0.1 and 0.6 ± 0.1 mas respectively (van Belle 1999). We cross-checked these diameters through SED model fits to photometric data available in the literature. Table 1 lists the calibrated visibility-squared measurements ($V^2_{total}$).

The measurements presented in this paper were taken over a narrow range of position angles (39–41º and 17–19º East of North) and projected baselines (83.9–84.9 m and 71.6–72.1 m) for the K/N and K/L measurements respectively.

We also carried out follow-up broadband observations of MWC 325 with the CHARA array in the *K*-band using the Classic beam-combiner and the E2-E1 (65.917m) baseline at position angle 65º on 16 July 2010. The calibrators used for these observations are HD188107, HD192343 and HD188953, and the adopted angular diameters are 0.186 ± 0.005, 0.206 ± 0.005 and 0.287 ± 0.004 mas, respectively. We used these observations (Table 1), in conjunction with our KI broadband (*K*-band) observations (Table 1) and *K*-band measurements in the literature (Table 2), to determine the inner-rim diameter and the inclination angle of the disk in the later part of Section 3.1.

| Wavelength (μm) | u (m) | v (m) | $V_{total}^2$ | $V_{total}^2$ error |
|---|---|---|---|---|
| Spectrally dispersed KIN measurements (UT 19 August 2008) | | | | |
| 2.004 | 54.548 | 64.593 | 0.1610 | 0.0419 |
| 2.102 | 54.548 | 64.593 | 0.1519 | 0.0414 |
| 2.169 | 54.548 | 64.593 | 0.1469 | 0.0411 |
| 2.288 | 54.548 | 64.593 | 0.1445 | 0.0409 |
| 2.377 | 54.548 | 64.593 | 0.1448 | 0.0407 |
| 8.209 | 54.548 | 64.593 | 0.4389 | 0.0073 |
| 8.765 | 54.548 | 64.593 | 0.4784 | 0.0104 |
| 9.186 | 54.548 | 64.593 | 0.5038 | 0.0068 |
| 9.730 | 54.548 | 64.593 | 0.5234 | 0.0083 |
| 10.270 | 54.548 | 64.593 | 0.5359 | 0.0108 |
| 10.726 | 54.548 | 64.593 | 0.5579 | 0.0098 |
| 11.232 | 54.548 | 64.593 | 0.5728 | 0.0142 |
| 11.741 | 54.548 | 64.593 | 0.5747 | 0.0219 |
| 12.223 | 54.548 | 64.593 | 0.6096 | 0.0390 |
| 12.713 | 54.548 | 64.593 | 0.6294 | 0.0859 |
| Spectrally dispersed KI measurements (UT 27 October 2009) | | | | |
| 2.004 | 22.261 | 68.356 | 0.1720 | 0.0408 |
| 2.102 | 22.261 | 68.356 | 0.1686 | 0.0400 |
| 2.169 | 22.261 | 68.356 | 0.1664 | 0.0400 |
| 2.288 | 22.261 | 68.356 | 0.1622 | 0.0400 |
| 2.377 | 22.261 | 68.356 | 0.1597 | 0.0400 |
| 3.487 | 22.014 | 68.367 | 0.2512 | 0.0400 |
| 3.568 | 22.014 | 68.367 | 0.2645 | 0.0400 |
| 3.629 | 22.014 | 68.367 | 0.2761 | 0.0401 |
| 3.734 | 22.014 | 68.367 | 0.2833 | 0.0401 |
| 3.808 | 22.014 | 68.367 | 0.2931 | 0.0401 |
| 3.846 | 22.014 | 68.367 | 0.3041 | 0.0401 |
| 3.978 | 22.014 | 68.367 | 0.3163 | 0.0401 |
| 4.029 | 22.014 | 68.367 | 0.3291 | 0.0402 |
| Broadband KI measurements (UT 19 August 2008) | | | | |
| 2.180 | 55.239 | 64.337 | 0.1435 | 0.0412 |
| Broadband KI measurements (UT 27 October 2009) | | | | |
| 2.180 | 22.261 | 68.356 | 0.1735 | 0.0412 |
| Broadband CHARA measurements (UT 16 July 2010) | | | | |
| 2.130 | 58.888 | 28.609 | 0.2303 | 0.0398 |

**Table 1: Calibrated $V_{total}^2$ is presented along with the wavelength, uv points and measurement errors.**

| UT Date | u (m) | V (m) | $V^2_{total}$ | $V^2_{total}$ error | Calibrators | Calibrator dia. (mas) |
|---|---|---|---|---|---|---|
| PTI Archive | | | | | | |
| 13 Oct 2003 | -77.805 | -24.805 | 0.1833 | 0.0313 | HD174160 | 0.38 ± 0.01 |
| | | | | | HD187923 | 0.80 ± 0.21 |
| | | | | | HD193556 | 0.73 ± 0.09 |
| 14 Oct 2003 | -48.412 | 65.934 | 0.1815 | 0.0333 | HD187923 | 0.80 ± 0.21 |
| | | | | | HD193556 | 0.73 ± 0.09 |
| KI Archive | | | | | | |
| 02 Jul 2007 | 40.935 | 67.021 | 0.1539 | 0.0313 | HD 183324 | 0.26 ± 0.01 |
| | | | | | HD 183385 | 0.64 ± 0.01 |
| | 32.564 | 67.769 | 0.1861 | 0.0318 | " | " |
| | 25.758 | 68.181 | 0.2046 | 0.0350 | " | " |
| | 20.092 | 68.438 | 0.2179 | 0.0344 | " | " |
| 16 Jul 2009 | 55.361 | 63.159 | 0.1040 | 0.0312 | HD 183926 | 0.35 ± 0.04 |
| | | | | | HD 187182 | 0.33 ± 0.04 |
| | | | | | HD 194244 | 0.30 ± 0.10 |
| | | | | | HD 190067 | 0.36 ± 0.01 |

**Table 2: Calibrated K-band (broadband) $V^2_{total}$ is presented along with uv points and measurement errors for KI and PTI archive data.**

As the K/N and K/L observations were taken at different epochs – separated by ~14 months – variability is an obvious concern when we combine these two sets of data. Since *K*-band measurements were taken on both occasions, we use this data to investigate possible variability. The mean *K*-band squared visibility was 0.14 ± 0.04 and 0.17 ± 0.04 for the epochs UT 19 August 2008 and UT 27 October 2009, respectively. These *K*-band $V^2$ measurements for the two epochs are comparable to within the one-sigma measurement errors. However, the baselines for these two epochs are somewhat different complicating direct comparison of these two measurements. Comparing these measurements with the *K*-band measurements in the literature (Table 2), taken during 2003 to 2010, show no evidence of size variability. Eisner et al. (2007), who were specifically looking for interferometric variability evidence in a small sample of YSOs, did not find any detectable size variations over time. We conclude, given the precision of our present observations, that there is no detectable size variability between these two epochs.

## 3. ANALYSIS

The measured $V^2_{total}$ includes contributions from the central star ($V^2_*$) and the circumstellar disk ($V^2_{disk}$). For the modeling analysis presented in this section, the visibility-squared of the circumstellar disk of

MWC 325 is obtained from the measured data by removing the contributions from the central star as follows. As the light from the central star and the circumstellar disk is incoherent, the total complex visibility is the sum of the complex visibilities of the star and the disk, i.e.,

$$\widehat{V}_{tot} = \frac{F_* \widehat{V}_* + F_{disk} \widehat{V}_{disk}}{F_* + F_{disk}}, \quad (2)$$

where $F_*$ and $F_{disk}$ are the stellar and disk fluxes, respectively. The squared modulus of this equation can be written in quadratic form as

$$V_{disk}^2 + 2rV_*V_{disk}\cos\theta + r^2 V_*^2 - (1+r)^2 V_{tot}^2 = 0, \quad (3)$$

where $r = F_*/F_{disk}$, $V_* = |\widetilde{V}_*|$, $V_{disk} = |\widetilde{V}_{disk}|$, and θ is the phase difference between the star and disk complex visibilities. The phase shift θ includes both the effective photocenter shift measured at the spatial frequency sampled by the interferometer, as well as the difference in the sign of the visibilities between the two components (for example, if one object were sampled on a negative lobe of its visibility function). If the disk and star are symmetric, and they are sampled on the main lobe of their visibility functions (i.e., they're reasonably compact), then θ = 2πB·δ/λ, where δ is the angular photocenter separation of the disk and star; this is the case in this paper. The solutions of this quadratic equation are

$$V_{disk} = -rV_*\cos\theta \pm [(1+r)^2 V_{tot}^2 - r^2 V_*^2 \sin^2\theta]^{\frac{1}{2}} \quad (4)$$

If the star and disk share the same effective photocenter, and they are both sampled on the main lobe of their visibility functions, then this can be written in terms of fringe amplitudes as

$$V_{disk}^2 = [(1+r)V_{tot} - rV_*]^2 \quad (5)$$

The stellar flux contribution in the *K, L* and *N* bands is estimated by fitting the Kurucz stellar atmospheric model (Kurucz 1970) for an A2-type star of solar metallicity ($T_{eff}$= 9250K, *log g* = 3.5) to dereddened BVR photometric fluxes (see below for the adopted interstellar extinction) and extrapolating the stellar atmospheric model at desired infrared wavelengths. The derived star-to-disk flux ratio, *r* is 0.2419, 0.0677 and 0.0086 in the *K, L*, and *N* bands. We use Equation 5 for the modeling work presented in Section 3.1 and 3.2, and for the initial modeling of Section 3.3. When the star and the disk do not share the same photo-center, and assuming symmetric, reasonably-compact objects, Equation 5 will overcorrect for the effect of the star, resulting in a lower estimated $V_{disk}^2$. For the final radiative transfer models presented in Section 3.3, we account for the photo-center effect (see Section 3.3 for the details).

The central star is assumed to be unresolved for our observations (i.e., $V_*^2$ = 1.0; the expected angular diameter of the central star is 0.12 mas for the assumed stellar radius and the distance, resulting in visibility-squared of > 0.999 for our instrument configurations). These central star corrections yield $V^2_{disk}$ values that are smaller than the total $V^2$ by 0.09 (~60%), 0.03 (~12%) and 0.003 (~0.6%) in *K*, *L* and *N* bands respectively. Such corrections are most important at the shortest wavelengths where the star is the brightest and the disk-to-star ratio is the smallest.

For our SED analysis, photometric measurements from the literature (2MASS All-Sky Catalog of Point Sources (Skrutskie et al., 2006); The Hipparcos and Tycho Catalogues (ESA 1997); IRAS catalogue of Point Sources, Version 2.0 (IPAC 1986); AKARI/IRC mid-IR all-sky Survey and AKARI/FIS All-Sky Survey Point Source Catalogues (ISAS/JAXA, 2010); Tannirkulam (2008)) are corrected for interstellar extinction using the extinction law of Cardelli et al. (1989). The extinction for MWC 325 in the *V*-band is assumed to be $A_v$ = 0.19 (Van den Ancker et al. 1998). We also used ISO-SWS (ISO Short-Wavelength Spectrometer) in the 2.36-4.1 μm wavelength region (Vandenbussche et al. 2002) and Spitzer spectra from the *Spitzer Space Telescope* in our SED modeling.

Our *K*-band $V^2$ measurements are consistent with earlier broadband *K* measurements, and provide additional spectrally-resolved information at different position angles enabling the determination of disk inclination angle. Our *L*-band and *N*-band measurements are unique and are also spectrally resolved.

### 3.1. SIMPLE WAVELENGTH-DEPENDENT GEOMETRICAL MODELS

In this section, we use our spectrally dispersed data within the *K*-, *L*- and *N*-band wavelength regions to investigate wavelength dependency of the source size and broadband (*K*-band) measurements to constrain the inclination angle of the disk. We chose three geometrical models, namely, uniform-disk, Gaussian distribution, and ring models to fit our spectrally dispersed measurements. The equations for these geometric models are given Ragland et al. (2009), where it was shown that the wavelength variation of the object size over the *K*- and *L*-bands was needed to fit the visibilities for MWC 419. However, we begin here for the current case, MWC 325, with the simplest possible wavelength-independent geometrical model and then motivate the need for a more complex model.

For wavelength-independent models, the derived uniform-disk, Gaussian, and ring angular sizes from simultaneous fits to our multi-wavelength measurements are 11.4 mas, 6.9 mas and 12.7 mas respectively. The corresponding reduced-chi-square $(\chi^2_R)$ values are 39, 28 and 45 respectively. As indicated by the very poor $\chi^2_R$ values, the wavelength-independent uniform disk, Gaussian, and ring models all fail to fit multi-wavelength visibilities in the *K, L,* and *N* bands.

Next we fit the data with wavelength-dependent sizes, taking advantage of the spectrally dispersed data across each of our bands. The resulting size-wavelength dependence for the uniform disk model in the 2–12 μm regions can be fit with a simple linear relationship, $\phi_{UD}(mas) = 4.594 + 0.736\lambda$, with a $\chi^2_R$ of 3.0. Increasing the complexity of the model even further, a log-quadratic fit of the form,

$\phi_{UD}(mas) = 3.431 + 5.392 \times \log(\lambda) + 3.297 \times [\log(\lambda)]^2$, to the derived apparent wavelength-dependent diameters (Figure 1; left) gives a $\chi_R^2$ of 0.3. The purpose of using an arbitrary function here is to get a rough knowledge of the size dependence as a function of wavelength. For comparisonn, size-wavelength dependence for a ring disk model is also shown in Figure 1 (right).

The models presented above assume face-on geometry for the disk consistent with the low *v sin(i)* measured for the star if the star and disk angular momentum axes are aligned. A disk with an inclination angle, *i* could have different properties, notably a larger inner disk radius, depending on the inclination angle and the position angle of the disk relative to the direction of the projected baseline, hence fringes.

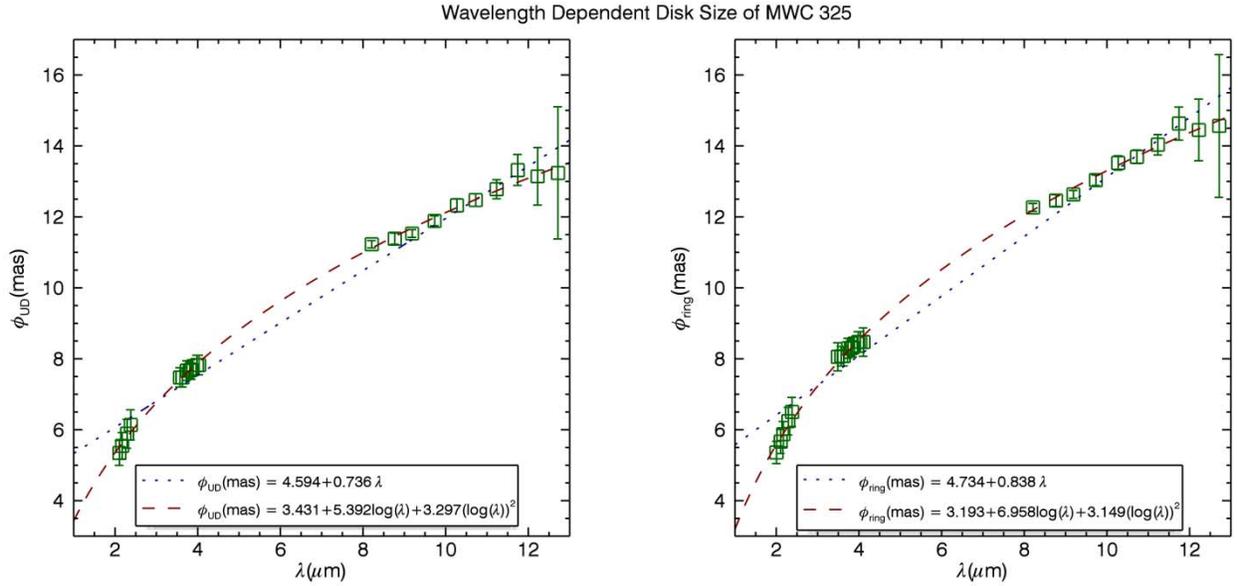

**Figure 1: Wavelength-dependent fits to the observed data. The estimated stellar contribution has been subtracted from the measurements. *Left:* Observed uniform-disk angular diameters as a function of wavelength are shown along with error bars. Linear (dotted, $\chi_R^2$ = 3.0) and log-quadratic (dashed, $\chi_R^2$ =0.3) fits to these diameters are also shown. *Right:* Same, except for ring model. The $\chi_R^2$ value for linear and log-quadratic model fits are 3.7 and 0.3 respectively.**

While multi-wavelength observations in the near- and mid-infrared wavelengths are effective in probing the extended disk, single wavelength measurements as a function of position angle could be used to determine the inclination angle of the disk. We used interferometric measurements in the *K*-band for this purpose to define the geometry of the inner-rim of the disk. We adopted an elliptical ring to represent the inner-rim and carried out model fit to our broadband *K*-band measurements from KI and CHARA (Table 1). We also included archived *K*-band interferometric measurements from KI (Keck Interferometer Archive, NExScI) on 02 July 2007 (PA: 31.42, 25.67, 21.92, 19.89º; Baseline: 78.53,

75.19, 73.38, 72.50m) and 16 July 2009 (PA: 41.24°; Baseline: 83.99m), and from PTI (Palomar Testbed Interferometer Archive, NExScI) on 13 Oct 2003 (PA: 72.32°; Baseline: 81.66m) and 14 Oct 2003 (PA: 143.71°; Baseline: 81.80m) in Table 2. The best-fit model is shown in Figure 2. The derived ring diameter, inclination angle and position angle of the major axis of the disk (East of North) are 5.85 ± 0.2 × 5.56 ± 0.2 mas, $18_{+30}^{-18}$ deg and $142_{+30}^{-30}$ deg respectively. In the following section we explore more detailed disk models to explain our observations.

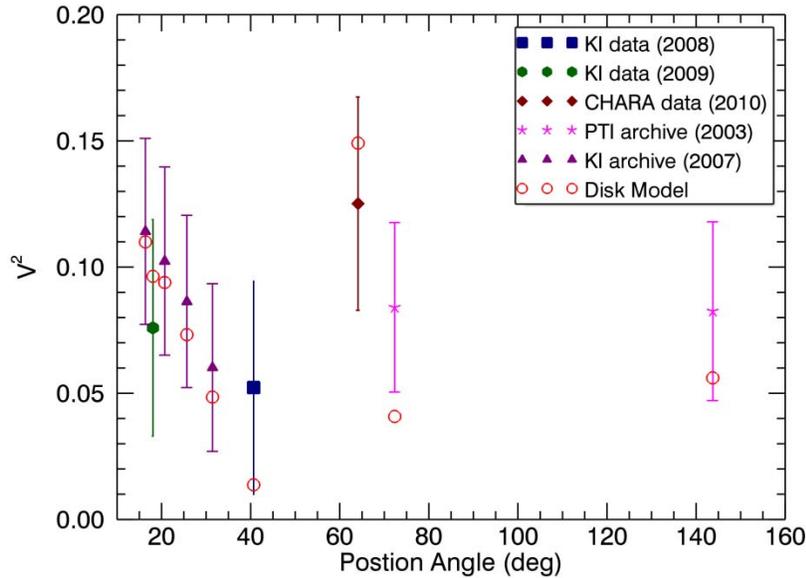

**Figure 2: Elliptical ring model fit to K-band interferometric data.**

### 3.2. GEOMETRICALLY THIN, OPTICALLY THICK DISK MODELS

Moving from simple geometric to more physically realistic models, we consider first a face-on accretion disk model (Hillenbrand et al. 1992) with a radial temperature distribution of the form $T(r) \propto r^{-\left(\frac{3}{4}\right)}$, where $r$ is the radial distance from the central star. The visibilities are computed by numerically summing the contributions from annular rings of infinitesimally small widths and weighting them by their respective flux contributions. We fit visibility and SED data simultaneously by treating the inner disk (hole) radius and the accretion rate as freely varying model parameters.

The best fit model parameters are given in Table 3 and the fits are shown in Figure 2. Even though the model itself is axisymmetric, since the effective projected baselines and position angles of the two epoch observations are different, two corresponding model curves are shown in Figure 2 through 4 (left panels). Notably, the classical accretion disk model fails to fit simultaneously the interferometric and the SED data. As illustrated in Figure 3, the fit to the interferometric data vastly under-predicts the amount of flux needed between 2–11 μm in the SED ($\chi^2_{IF}=2.6, \chi^2_{SED}=9.8 \,\&\, \chi^2_{Total}=8.0$). Specifically, the derived inner disk temperature is 628 K while the SED requires hotter dust. The derived inner angular diameter of the face-on disk is 4.02 ± 0.04 mas. The fit to the interferometric data points could be improved ($\chi^2_{IF}=1.1$) by changing the inclination angle of the accretion disk to 70° and the mass accretion rate to 8 x $10^{-7}$ $M_\odot$ yr$^{-1}$. However, such a model highly underestimates SED flux $\chi^2_{SED}=28.3 \,\&\, \chi^2_{Total}=21.5$, particularly in the near- to mid-infrared from about 1 to 10 microns. $\chi^2_{Total}$ is computed combining both squared-visibilities and SED data with their measurements errors.

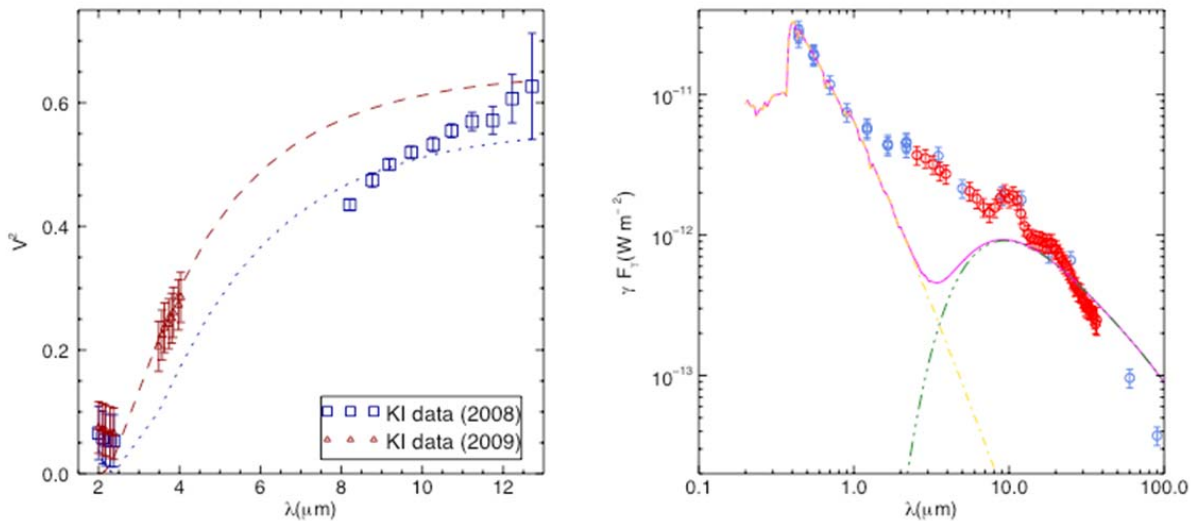

**Figure 3: Face-on classical accretion disk model fits.** *Left:* **Interferometric data points are shown, in square and triangle symbols for observations taken at UT 19 August 2008 and UT 27 October 2009 respectively, along with error bars. The estimated stellar contribution has been subtracted from the measurements. The dotted and dashed lines represent an accretion disk model for the baseline orientation of the two epochs. The inner radius of the disk (hole size) is treated as a free model parameter and the outer radius of the disk is fixed at 100 AU.** *Right:* **Photometric data taken from the literature (see text) are shown along with the SED model for the same accretion disk model. The dash-dot, dash-dot-dot-dot, and solid lines are model SEDs of the star, disk, and star-plus-disk respectively.**

One way to get hotter dust with only small changes to the size of the inner hole is to steepen the disk temperature gradient. We thus fit our data with a power-law temperature gradient model of functional form $T(r) \propto r^{-\alpha}$, where $r$ is the radial distance from the central star and $\alpha$ is the power-law parameter (0.75 for the classical accretion disk). The temperature of the inner edge of the dust disk is fixed at 1500K, corresponding to a notional dust destruction radius. The inner radius and $\alpha$ are treated as free model parameters. The inclination angle of the disk is fixed to zero in the case of face-on power-law disk, but varied as a free parameter in the case of inclined power-law disk.

The resultant model fit (in the case of inclined power-law disk) is shown in Figure 4 and the model parameters are given in Table 3. The derived angular diameter of the inner dust edge along the major axis and the power-law exponent are 3.86 ± 0.10 mas and 1.12 ± 0.02 for the face-on disk, and 4.47 ± 0.20 mas and 1.26 ± 0.05 for the inclined-disk respectively. The inclination angle of the disk is ~ 72 ± 4° and the position angle of major-axis of the disk is ~ 49 ± 4°. While our model fit to the interferometric and SED data is improved over the classical accretion disk model, the SED fit under-predicts the 1–3 µm and 10–60 µm flux, and over-predicts the 3–9 µm flux. This parameterized disk model is meant to approximate the emission near the "inner dust rim" disk component, which dominates the infrared emission, and which for this type of star has been found to often be required in order to explain the spatially resolved IR observations (see e.g. the recent review by Dullemond & Monnier 2010). The seemingly steep temperature gradients derived from these power-law models are discussed in Section 4.

As we mentioned before, the distance to MWC 325 is uncertain. We adopted a distance of 340 pc as the blackbody model fits to BVRI photometric data favors this value. If we were to use 767 pc (Montesinos et al. 2009), the inner radius of the disk would have a larger value for the accretion disk and the power-law models. The inner radius of the face-on accretion disk, face-on power-law and inclined power-law models are 1.51 ± 0.01 AU, 1.48 ± 0.02 AU and 1.71 ± 0.05 AU respectively. While the other parameters of the power-law models are not affected, some parameters of the face-on accretion disk model changed; the temperature of the inner edge decreased to 645 K and the mass accretion rate increased to 4 x $10^{-5}$ $M_\odot$ yr$^{-1}$. The parameters of the disk models presented in this section should be treated with some caution especially the inclination angle (and hence position angle) and mass accretion rate that are arbitrarily used to scale the disk flux in order to match the overall flux level of the SED data. However, they do not fit the SED data satisfactorily. In the following section, we present a more complex physical model based on radiative transfer calculations to improve fit to our interferometric measurements and SED data.

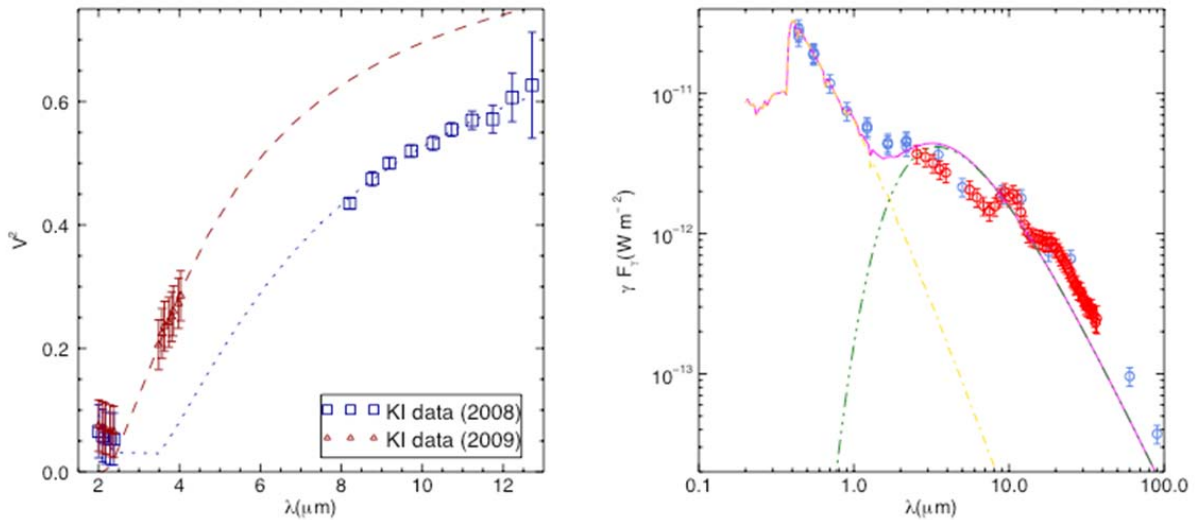

**Figure 4: Power-law disk model fits.** *Left:* Symbols are as in Figure 2. The dotted and dashed lines represent a power-law temperature gradient disk model for the baseline orientation of the two epochs. The power-law parameter, the inner disk radius, and the inclination angle and the position angle of the disk are free parameters. The dust temperature at the inner edge, and the outer radius, are fixed at 1500 K and 100AU respectively. *Right:* Symbols as in Figure 3 are shown along with the SED model for the same power-law model. The dash-dot, dash-dot-dot-dot, and solid lines are model SEDs of the star, disk, and star-plus-disk respectively.

| Model Parameters | Face-on Accretion disk | Face-on Power-law disk | Inclined Power-law disk |
|---|---|---|---|
| $R_{in}$ (AU) | $0.68 \pm 0.01$ | $0.66 \pm 0.02$ | $0.76 \pm 0.05$ |
| $T_{in}$ (K) | 628 | 1500 (fixed) | 1500 (fixed) |
| Radial power-law exponent, $\alpha$ | 0.75 (fixed) | $1.12 \pm 0.02$ | $1.26 \pm 0.05$ |
| Disk inclination, $i$ | $0°$ (fixed) | $0°$ (fixed) | $72 \pm 4°$ |
| Position angle of the major-axis | - | - | $49 \pm 4°$ |
| $M_{acc}$ ($M_\odot$ yr$^{-1}$) | $3 \times 10^{-6}$ | - | - |
| $\chi^2_{IF}$ | 2.6 | 1.3 | 0.6 |
| $\chi^2_{SED}$ | 9.8 | 126.0 | 5.9 |
| $\chi^2_{Total}$ | 8.0 | 94.9 | 4.6 |

**Table 3:** Derived parameters for the disk models presented in Section 3.2. The entries $\chi^2_{IF}$, $\chi^2_{SED}$, and $\chi^2_{Total}$ refer to the $\chi^2_R$ for interferometric data, SED data, and the combined set of interferometric and SED data respectively.

### 3.3. RADIATIVE TRANSFER MODELS

We have also carried out two-dimensional radiative transfer modeling of the dusty disk of MWC 325 through Monte Carlo simulations (Ohnaka et al. 2006) in order to explain our interferometric measurements. In this radiative transfer calculation, photon packets randomly released from the stellar surface are subjected to absorption and scattering by multiple species of dust grains in the circumstellar disk (we assumed isotropic scattering for simplicity). The dust temperature of each grain species is computed using the method of Bjorkman & Wood (2001). The outputs of the Monte Carlo code are the SED viewed from arbitrary inclination angles, the temperature of each grain species, and the monochromatic mean intensity at each cell position in the dust disk. Using these resultant mean intensity and dust temperatures, monochromatic images viewed from any arbitrary angle are computed through ray tracing.

We used a Kurucz stellar atmospheric model (Kurucz 1970) for an A2-type star of solar metallicity ($T_{eff}$ = 9250K, $\log g$ = 3.5) as our input stellar spectrum for our radiative transfer calculations. We assumed that the disk consists of a mixture of graphite and silicate with equal fractional abundances and used the optical properties presented by Draine & Lee (1984). For the modeling of MWC 325, we adopted representative "sub-micron" (0.1 μm) and "micron" (2 μm) radius grains. We computed models with different grain size distributions, for example, models only with the sub-micron sized grains or micron-sized grains, as well as models with mm-sized grains. However, it turned out that a two-grain model consists of 0.1 and 2 μm grain is needed to reproduce the observed SED and interferometric data.

The dust density distribution in the circumstellar disk is characterized by the standard flared disk geometry given by

$$\rho_i(r,z) \propto \left(\frac{r}{r_0}\right)^{-p} \exp\left[-\frac{1}{2}\left(\frac{z}{h_i(r)}\right)^2\right] \qquad (6)$$

$$h_i(r) = h_{i,0}\left(\frac{r}{r_0}\right)^q \qquad (7)$$

where $r$ is the radial distance in the equatorial plane, $z$ is the height from the equatorial plane, and $h_{i,0}$ is the scale height of the i$^{th}$ grain species at some reference radius $r_0$. In the region with $r$ smaller than the inner boundary radius, dust was assumed to be absent.

In order to allow for the vertical dust segregation that has been inferred (Tannirkulam et al. 2007), we treated the scale heights of the small and large grains as free parameters. The other free parameters are the disk's inner boundary radius of the dust grains (assumed to be the same for both grain species) and the optical depth of dust grains (equivalent to specifying the mass of dust grains).

The computation is CPU-intensive and performed in a parallel computing environment, simultaneously running on three Sun Fire V440 servers (each one powered by four 1.593 GHz UltraSPARC IIIi processors), taking about two days of CPU time to generate a single model with adequate SNR, spatial resolution and field-of-view. We used $2\times10^7$ photon packets for most of our simulations for stable results with adequate SNR. A library of model images at the wavelengths of interest and the corresponding model SEDs are generated for a range of model parameters. Model $V^2$ values are computed from these images and compared with measured $V^2$ data by varying the position angle of the disk through non-linear least squares fitting. The choice of distance has no impact on our radiative transfer modeling since the raw images are in units of stellar radius and are scaled to fit the SED data. The scaling of these images gives a distance of 321 pc to MWC 325 within roughly 5% of the adopted 340 pc value.

Our attempt to fit simultaneously interferometric and SED data with a flared-disk model (q ≥ 1) was not successful. In these models, the outer region of the disk is efficiently warmed up so that the 10 micron silicate emission originates in the outer region of the disk. This makes the object appear larger in the 10 micron silicate feature than at 8 or 13 micron. However, as shown in Fig. 1, the size of the object increases monotonically from 8 to 13 micron without regard to the presence of the 10 micron feature, whether in emission or absorption. The situation is different from that typical of studies of gas in circumstellar disks in which the visibilities across H and CO features change relative to the continuum, allowing conclusions on the relative spatial distribution of gas and dust. We conclude that flared disk models with q ≥ 1 cannot explain the observed N-band visibilities.

Subsequently we explored the parametric space, q < 1 and the best fit provides q = 0.875, suggesting that the disk is fairly flat with little or no flaring. Comparison between our observations and the best-fit model is shown in Fig. 4. The best-fit to the data can be obtained by a disk model viewed from an intermediate inclination angle of 45 degrees with a dust sublimation radius of 1.26 AU.

As mentioned before, if the photocenters of the star and disk are not the same, the use of Equation 5 incorrectly estimates the squared visibility of the disk. We use Equation 5 for the initial models. Then we derive fringe phase information from the model images to compute an improved set of squared visibilities for the disk using Equation 4. We iterate this procedure a few times until it converges. The maximum correction required for the photo-center effect was 0.035 in squared visibility - well within the one-sigma measurement errors. The assumed stellar parameters are given in Table 4 and the characteristics of the flat-disk are given in Table 5. Figure 5 (left) shows that the interferometric data spanning from 2 to 13 µm are well reproduced by this model. Figure 5 (right) shows that the near-IR part of the observed SED, as well as the mid-IR excess including the 10 µm silicate emission, is reasonably reproduced, although the model predicts values of the far-IR 60 and 90 µm flux well below the observations. However, the model far-IR values could be considered as consistent with the measurements given the broad bandwidth of these far-IR measurements (shown in Figures 5 & 6 as horizontal error bars). This model gives a $\chi_R^2$ value of 1.5 for the combined set of interferometric and SED data. However, if we restrict to only 2-13µm region of the SED fit, we get a $\chi_R^2(2-13\mu m)$ value of 1.0. In this model with little or no flaring, which slightly shadows the intermediate distance regions, the outer region of the disk is not efficiently warmed up. Therefore, the 10 µm silicate emission is confined in the inner region which prevents the object from appearing larger at 10 µm than at 8 or 13 µm. The increase in the object size from 8 to 13 µm simply results from the emission at longer wavelengths originating from the larger distances from the star.

In order to improve the fit to far-IR SED data further, we also developed a model that flares only in the outer regions meaning that the inner region is vertically thick enough to shadow a significant part of the disk except the outer regions, resulting in flaring at these regions. The resultant model fit is shown in Figure 6 and the disk parameters are given in Table 5. This model, which is the best-fit model of all those considered, gives a $\chi_R^2$ value of 1.1 for the combined set of interferometric and SED data. Even if we consider only 2-13µm region of the SED fit, we get a similar value, $\chi_R^2(2-13\mu m) \sim 1.1$. The model SED corresponding to the weakly-shadowed model compares well with the far-IR SED data, but not the near-IR SED. The errors of the model parameters given in Table 5 refer to the $\chi_R^2+1$ location of the chi-square space. These errors can be underestimated in the case of two or more parameters being degenerate. The scale height of the large grains is found to be about 80% of that of the small grains, somewhat larger than the value of 60% used by Tannirkulam et al. (2007). The power-law exponents *p* and *q* (Eq. 2) in the inner regions of the disk are 1.8 and 0.825 respectively providing a surface density distribution of $\Sigma \propto r^{-0.975}$. These parameters for the outer flaring regions are 1.8 and 1.125 respectively providing a surface density distribution of $\Sigma \propto r^{-0.675}$. The corresponding model images at 2.17, 3.81, 8.21, 9.73 and 12.71 μm for this best-fit weakly-shadowed disk model are given in Figure 7. These images suggest that the *K*-band data are dominated by the emission from the inner rim, while the extended emission becomes more and more prominent from the *L* to *N* band. The intensity profiles of the disk along the major and minor axes are shown in Figure 8.

The radial distribution of grain temperature and grain mass density along the mid-plane for our best-fit weakly-shadowed disk model is given in Figure 9. The vertical distance of the $\tau_\lambda=1$ surface at 2.17, 3.81, 8.21, 9.73 and 12.71 μm of this model is also given in Figure 9. The height of $\tau_\lambda=1$ surfaces reflect the wavelength dependence of the opacity. The $\tau_\lambda=1$ surface for 3.81 micron is higher than that for 2.17 micron, because the scattering coefficient of the 2 micron-sized grains shows a peak at 3-4 micron. Likewise, the surface for 9.73 micron is higher than those for 8.21 and 12.71 micron because of the 10 micron silicate feature. The vertical distance of $\tau_\lambda=1$ surface goes down at large radii. This is because there are two competing factors in determining this surface: (1) density decrease in the radial direction and (2) scale height increases at larger radii. At large radii, the former factor is greater than the latter, leading to the downturn of the $\tau_\lambda=1$ surface. Moreover, the plot does not go all the way to the outer radius (= 100 AU), because the optical depth measured from the vertical infinity does not reach 1 before the mid-plane at radii larger than some value, which depends on wavelength. The peak value of the $\tau_\lambda=1$ surface is in the range 0.25-0.45 AU. As Fig. 11 (upper panels) show, the tempertures at these vertical distances are not different from that of the mid-plane and is more or less same in the 2-13 μm region. A representative temperature profile at 3.81 μm (where the vertical distance of the $\tau_\lambda=1$ surface is maximum) of the best-fit weakly-shadowed disk model is shown in Figure 9 (bottom right).

In a flared disk, the stellar and any other high energy flux is absorbed and re-radiated in the upper disk atmosphere layers. Mid-plane heating is possible either via re-radiation or directly if stellar photons entering from the top of the disk atmosphere can propagate downward to the mid-plane. However, in the case of MWC 325, flaring becomes important only beyond 13 AU from the star (see Figure 9, upper left panel). While the argument above about the stellar flux possibly heating the mid-plane of the disk is appropriate for the flared regions, where there is a geometrically large area to intercept radiation from the star, at small disk radii this is not the case. The radius of the star is only 0.017 AU in size compared to the 0.1 AU scale height of the disk at 1.26 AU. Thus the star can only effectively illuminate the inner radius of the disk and it of course illuminates some of the disk but that material is of low density being a few scale heights from the mid-plane and hence can only weakly heat the mid-plane since it is low density material that is re-radiating the star light. The material has to be something greater than about 4 degrees (i.e. $\tan^{-1}((0.1-0.017)/1.26)$) in angle from the star off the mid-plane to get mostly pure stellar radiation.

We derive a total dust mass of $9.9 \times 10^{-8}$ $M_\odot$ for sub-micron grains and $1.5 \times 10^{-6}$ $M_\odot$ for the micron grains, assuming a bulk density of 3.5 gm cm$^{-3}$ and 2.24 gm cm$^{-3}$ for the silicate and graphite grains

(Zubko et al. 2004) and hence 2.87 gm cm$^{-3}$ for the silicate-graphite (50:50 ratio) mixture. Thus the derived total dust-mass of the sub-micron and micron sized grains in the disk of MWC 325 is $1.6 \times 10^{-6}$ $M_\odot$.

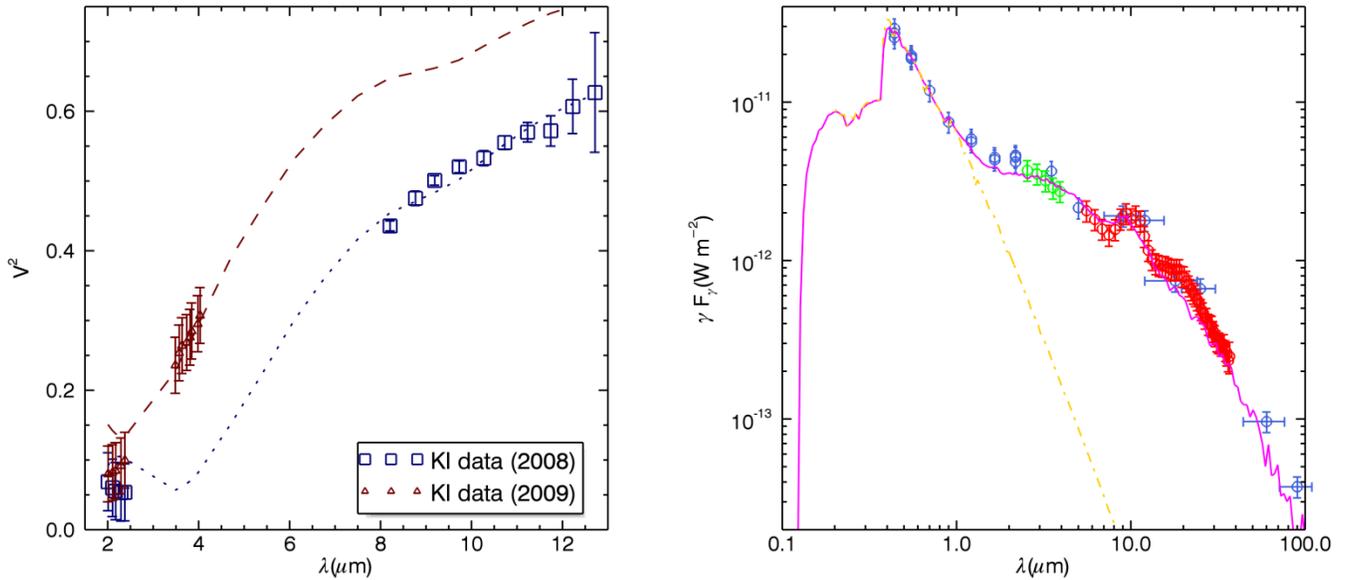

**Figure 5: Radiative transfer model fits.** *Left:* **Multi-wavelength $V^2$ measurements and model $V^2$ Monte-Carlo simulations (see text for details). The dotted and dashed lines represent visibilities predicted by the 2-D radiative transfer model of a flat disk for the baseline orientation of the two epochs shown along with symbols as in Figure 3.** *Right:* **The dash-dot and solid lines are model SEDs of the star and star-plus-disk respectively while symbols are as in Figure 3. The bandwidth (FWHM) of the IRAS and AKARI photometric data are shown as horizontal error bars.**

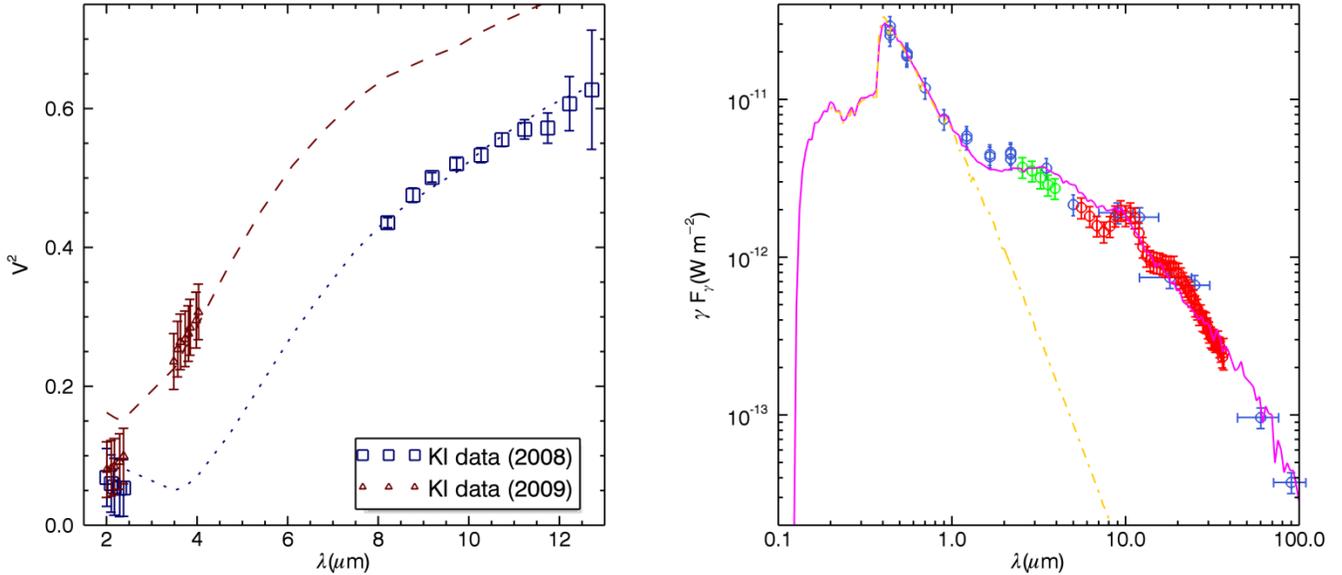

**Figure 6: Radiative transfer model fits.** *Left:* **Multi-wavelength $V^2$ measurements and model $V^2$ Monte-Carlo simulations (see text for details). The dotted and dashed lines represent visibilities predicted by the 2-D radiative transfer model of a weakly-shadowed disk for the baseline orientation of the two epochs. Symbols are as in Figure 3.**

*Right:* The dash-dot and solid lines are model SEDs of the star and star-plus-disk respectively while symbols are as in Figure 3. The bandwidth of the 60 and 90 μm measurements are shown as horizontal error bar.

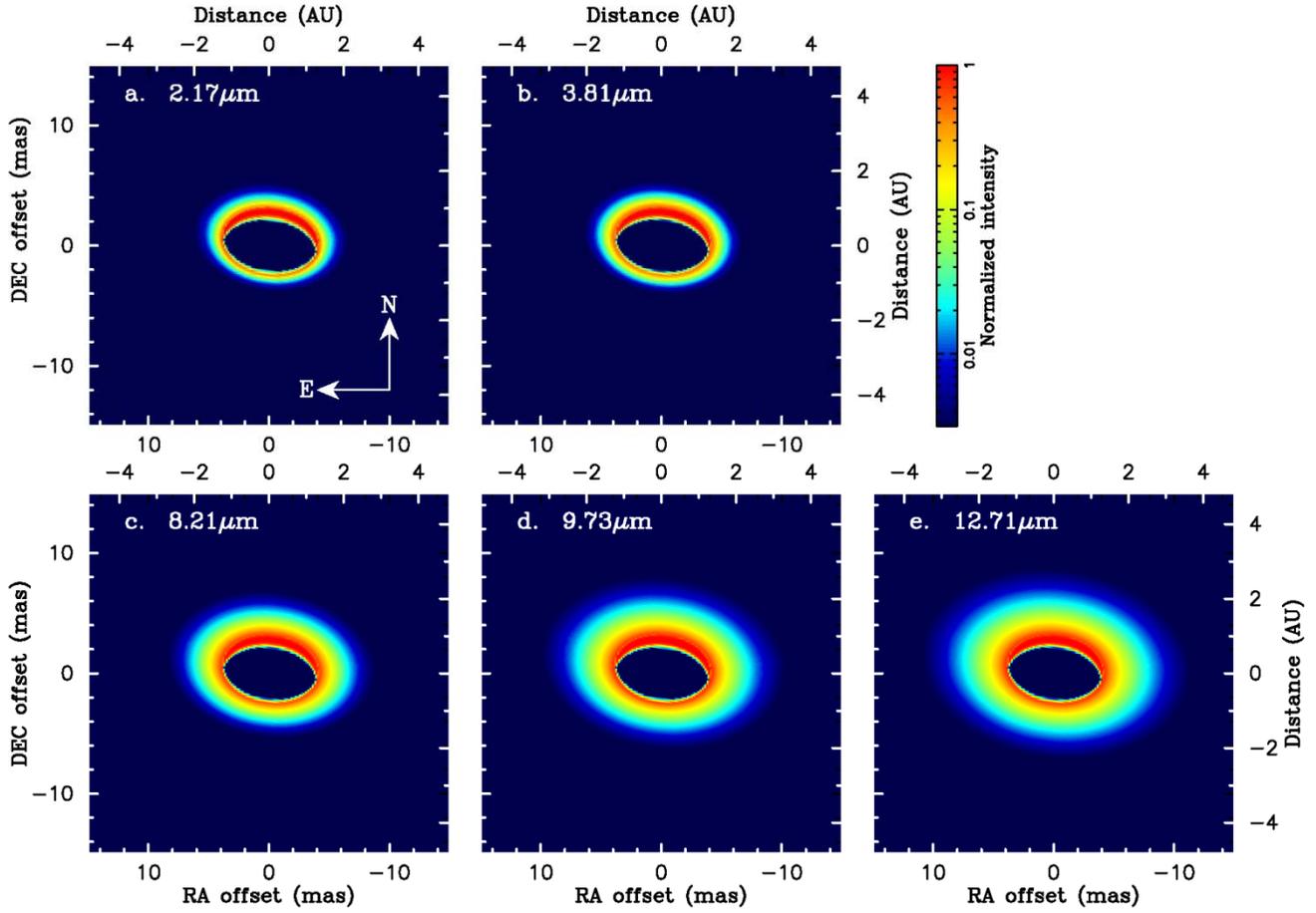

**Figure 7:** Model images in logarithmic scale. Weakly-shadowed disk model images generated through radiative transfer modeling of multi-wavelength interferometric measurements and SED data are shown at 2.17, 3.81, 8.21, 9.73 and 12.71 μm (see the text for details). The disk parameters are given in Table 5 (right column).

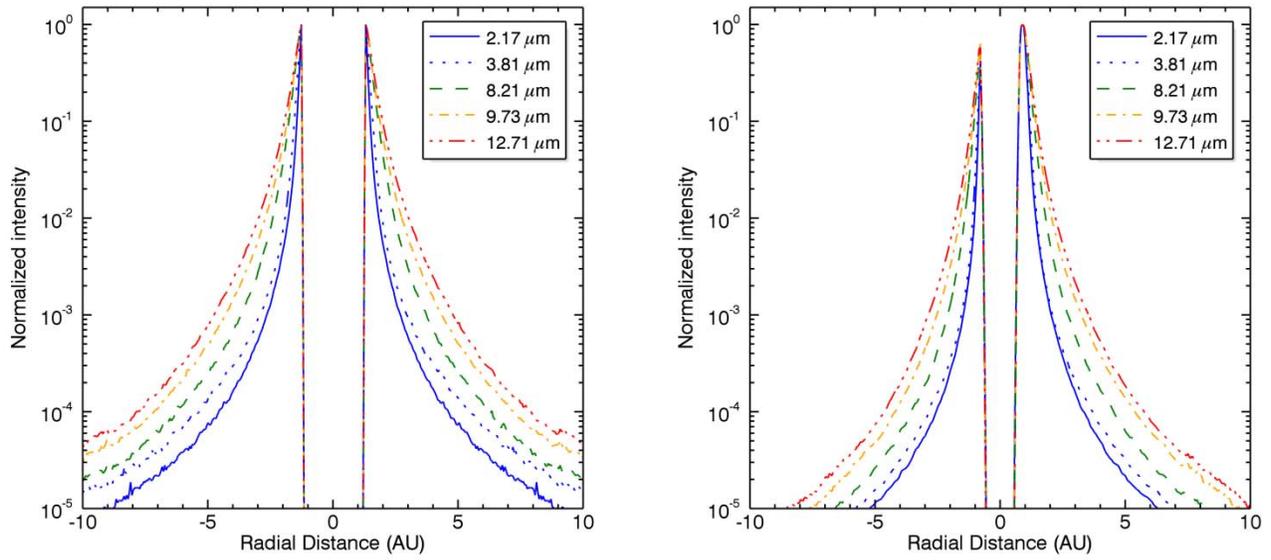

**Figure 8:** Left: Intensity profile of the best-fit weakly-shadowed disk model shown in Figure 7 at 2.17, 3.81, 8.21, 9.73 and 12.71 μm along the major axis. Right: same as that of the left Figure but for the minor axis.

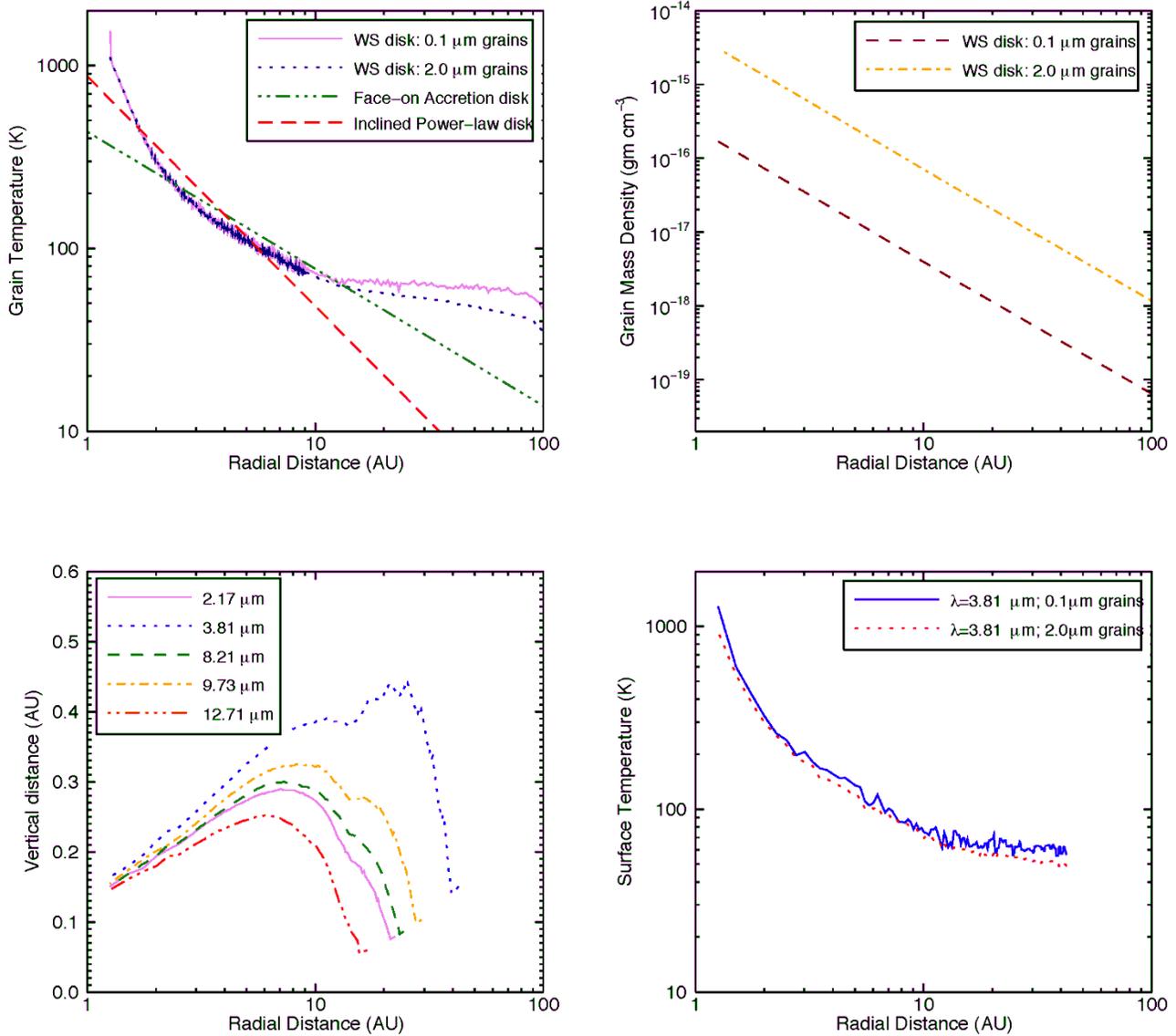

Figure 9: Top-left: Radial distribution of grain temperature of our best fit weakly-shadowed (WS) disk model. Solid and dotted lines represents radial temperature profile of 0.1 μm and 2 μm grains, respectively, along the mid-plane of the disk. Also shown for comparison are the temperature profiles of the best-fit face-on accretion disk (dash-dot-dot-dot line) and inclined power-law disk (long dashes line) models. To-right: Radial distribution of grain mass density along the mid-plane of the disk for both 0.1 μm (solid line) and 2 μm grains (dotted line). Bottom-left: Vertical distance of the tau=1 surface at 2.17, 3.81, 8.21, 9.73 and 12.71 μm (see the text for details). Bottom-right: Surface temperature of the disk at 2.17 μm for both grain sizes is shown. The temperature profiles for other wavelengths look similar and are not shown here for clarity (see the text for details).

| Stellar Parameters | Value |
|---|---|
| Spectral type | A2 IVev |
| Stellar luminosity | 83 $L_\odot$ |
| Stellar effective temperature | 9250 K |
| Stellar radius | 3.6 $R_\odot$ |
| Stellar mass | 2.85 $M_\odot$ |
| Distance | 340 pc |

**Table 4: Assumed stellar parameters**

| Model Parameters | Flat disk | Weakly-shadowed disk |
|---|---|---|
| Radial power-law exponent, $p$ | 1.8 ± 0.8 | 1.8 ± 0.8 |
| Vertical power-law exponent, $q$ | 0.875 ± 0.13 | 0.825 ± 0.13 |
| Scale height ($h_0$) of sub-micron grains at 1.26AU | 0.10 ± 0.02 AU | 0.10 ± 0.02 AU |
| Scale height ($h_0$) of micron grains at 1.26 AU | 0.08 ± 0.04 AU | 0.08 ± 0.04 AU |
| Inner boundary radius for both grains | $1.26^{+0.03}_{-0.08}$ AU | $1.26^{+0.03}_{-0.08}$ AU |
| Inner boundary radius for Flaring (fixed) | NA | 13.4 AU |
| Vertical power-law exponent, $q_{outer}$ (fixed) | NA | 1.25 |
| Inner boundary dust temperature of 0.1 μm grains | 1604 K | 1537 K |
| Inner boundary dust temperature of 2 μm grains | 1094 K | 1065 K |
| Representative grain sizes (fixed) | 0.1 μm & 2.0 μm | 0.1 μm & 2.0 μm |
| Outer boundary radius (fixed) | 100 AU | 100 AU |
| Disk inclination | $45^{+25}_{-20}$ degrees | $45^{+25}_{-20}$ degrees |
| Position angle of the disk major axis | $90^{+40}_{-40}$ degrees | $85^{+40}_{-40}$ degrees |
| Radial optical depth of the grains at 0.55μm ($\tau_v$) | $200^{+800}_{-125}$ | $200^{+800}_{-125}$ |
| Total dust mass of the disk | $1.1 \times 10^{-6} M_\odot$ | $1.6 \times 10^{-6} M_\odot$ |
| $\chi^2_{IF}$, $\chi^2_{SED}$ & $\chi^2_{Total}$ | 1.24, 1.55 & 1.47 | 1.03, 1.14 & 1.13 |
| $\chi^2_{IF}$, $\chi^2_{SED}$(2-13μm) & $\chi^2_{Total}$(2-13μm) | 1.24, 0.59 & 0.98 | 1.03, 1.11 & 1.06 |

**Table 5: Derived parameters for the circumstellar disk.** The entries $\chi^2_{IF}$, $\chi^2_{SED}$, and $\chi^2_{Total}$ refer to the $\chi^2_R$ for interferometric data, SED data, and the combined set of interferometric and SED data respectively. $\chi^2_{SED}$(2-13μm) and $\chi^2_{Total}$(2-13μm) refer to the $\chi^2_R$ for SED data, and the combined set of interferometric and SED data in the 2-13μm region.

## 4. DISCUSSION

Our new spectrally dispersed *K*-, *L*-, and *N*-band KI observations provide powerful new constraints on the physical structure of the material surrounding the young Herbig Ae star MWC 325. Wavelength-dependent uniform-disk sizes of our measurements have a steep slope (i.e. size increases with wavelength) confirming that the 2–12 μm emission region is extended with strong radial temperature dependence.

For MWC 325, the derived uniform-disk diameter at 10 μm ($\phi_{10\mu m}$ = 12.3 ± 0.2 mas) is about a factor of 2.2 larger than the diameter at 2.2 μm ($\phi_{2.2\mu m}$ = 5.7 ± 0.3 mas), and the derived mean uniform-disk diameter at 3.7 μm ($\phi_{3.7\mu m}$ = 7.7 ± 0.2 mas) is about a factor of 1.4 larger than the diameter at 2.2 μm. The derived *K*-band uniform-disk angular diameter is consistent with the reported *K*-band value of $5.57^{+0.04}_{-0.04}$ mas by Eisner et al. (2004). Monnier et al. (2006) reported a ring diameter of 3.48 ± 0.4 mas in the *H*-band. Eisner et al. (2009) used a dust plus gas model to fit their spectrally dispersed (R ~ 230) KI observations in the *K*-band. These authors report a dust ring diameter of 4.28 ± 0.07 mas at 1106 ± 10 K, and an inner gas diameter of 0.97 ± 0.07 mas at 3115 K. Our modeling shows that the observed SED and KIN data can be explained without hot emission from inside the dust sublimation radius. This can indeed mean the absence of the hot emission found in some Herbig Be stars (due to hot gas or highly refractory dust). However, since we did not test a model with such hot emission, we cannot entirely conclude on this.

The disk inclination angle of $45^{+25}_{-20}$ degrees derived from our radiative transfer calculations is consistent with that of the inner-rim of $18^{-18}_{+30}$ degrees derived from our elliptical ring model fit to broadband *K*-band measurements within the measurement accuracy. Moreover, the best-fit radiative transfer model also fits *K*-band measurements well ($\chi^2_{IF}$ = 1.3). Earlier, Isella et al. (2006) report disk inclination angle in the range 40 – 65 degrees using *H*- and *K*-band interferometry data available in the literature, which is somewhat larger than our value. And their inner boundary radius of 0.7 AU is about a factor of two smaller than our value of 1.26. While our analysis incorporates sub-micron (0.1 μm) and micron (2 μm) radius grains to fit the interferometric and SED data simultaneously, Isella et al. (2006) used 0.3 μm and ≥1.2 μm radius grains to fit simultaneously near-infrared interferometric data and optical and near-infrared SED data (far-infrared SED data was not included in their analysis). Kraus et al. (2008) used similar small (0.005–1 μm) and large (1–1000 μm) radius grains to simultaneously fit broad wavelength range of interferometric and SED data of the Herbig Be star, MWC 147. We suggest that the grains in the disk of MWC 325 haven't grown to millimeter-sized grains. Perhaps the luminosity of the central star plays a major role in promoting the grain growth process.

The radial distribution of grain temperature of MWC 325 along the $\tau_\lambda=1$ surface of the disk (Figure 9) in the 1-2 AU region is steeper than that of classic disk with a power-law exponent of -0.75 and the temperature gradient becomes shallower at larger radial distances. For deriving power-law index of -0.75, the disk is assumed to be directly irradiated by the star. However, this is not the case for the material in the $\tau_\lambda=1$ surface except at the inner-rim. Figure 10 shows the optical depth at 0.55 micron in the mid-plane measured from the inner rim. The figure shows that the optical depth to the region even slightly behind the inner rim is already higher than 1, and therefore, the $\tau_\lambda=1$ surface of the disk beyond the inner-rim region is shielded from the stellar radiation, resulting in reduced grain temperature (see the temperatures around the mid-plane at three representative radial positions in Figure 11) and hence steeper temperature gradient.

Earlier multi-band interferometry has shown somewhat similar temperature profile for two Herbig Ae disks, namely, MWC 275 and AB Aur (Tannirkulam et al. 2008a) and to lesser extent for the Herbig Be disk MWC 147 (Kraus et al. 2008a). Recent disk models include an inner rim at the boundary between the dust-free inner disk and the main disk that is vertically "puffed up" relative to the dusty material at only slightly larger radii. Modifications to this simple geometry and with some hot refractory dust or gaseous material inside the inner rim and a blunt (rather than sharp) dust boundary can have just the effect we see of an apparent size gradient with wavelength and an accompanying steep temperature profile (see e.g. the recent review by Dullemond & Monnier 2010). Radial sorting of grain sizes may also play a role in the effects that we see.

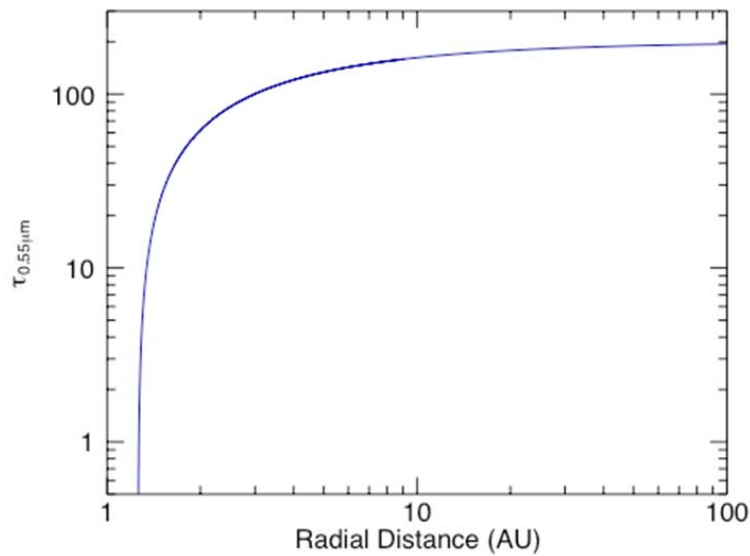

**Figure 10: Radial distribution of optical depth along the mid-plane of the best-fit weakly-shadowed disk, i.e., at latitude 0 deg. As can be seen, the disk is optically thick at the inner regions along the mid-plane shielding the outer regions from the stellar radiation.**

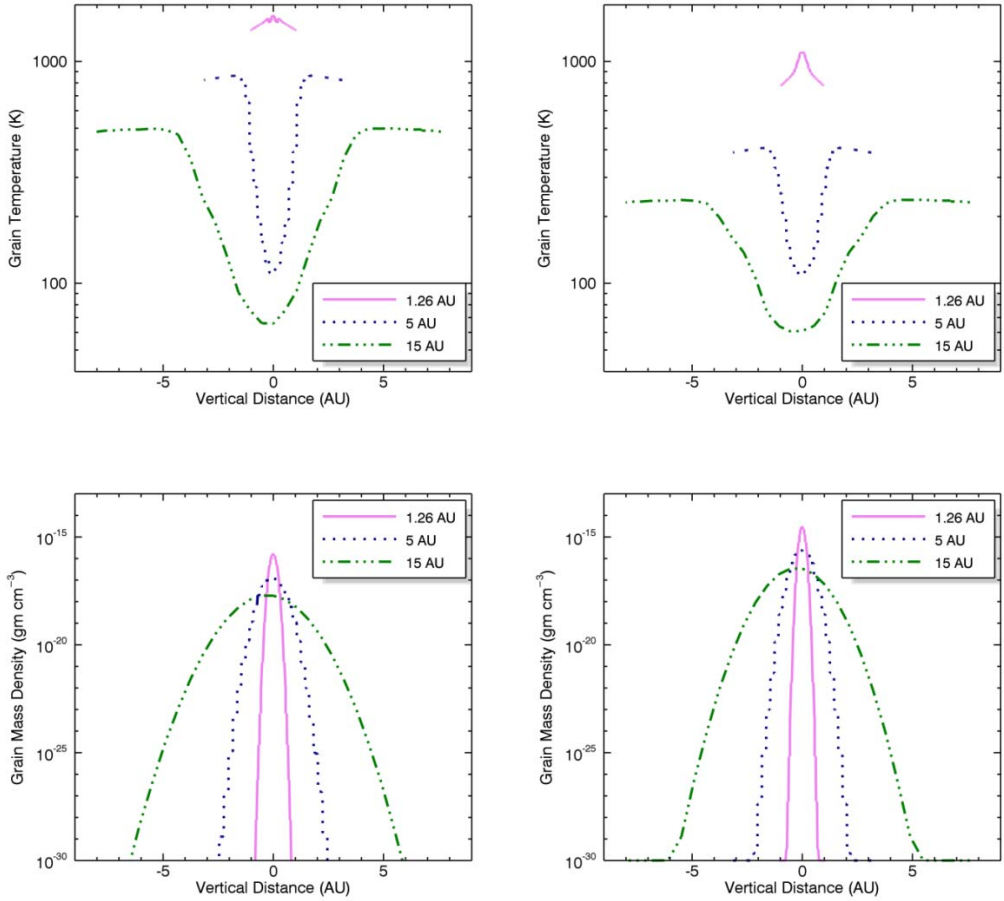

**Figure 11:** Vertical temperature and mass density distributions of our best fit radiative transfer model of a weakly-shadowed disk. Top Left: Solid, dotted and dashed lines represents vertical temperature profile of 0.1 μm grains at three radial distances, namely, 1.26AU (inner radius), 5 AU & 15 AU respectively. Top Right: Same as the top left plot for 2 μm grains. Bottom Left: Solid, dotted and dashed lines represents vertical mass density profile of 0.1 μm grains at three radial distances, namely, 1.26AU (inner radius), 5 AU & 15 AU respectively. Bottom Right: Same as the bottom left plot for 2 μm grains.

We assumed a grain composition consists of a mixture of graphite and silicate with equal fractional abundances for the radiative transfer models presented in this paper. We also attempted a grain composition of Silicate and amorphous carbon with equal abundances. We find that such a grain composition generates weak 10μm feature. The reason is that graphite has steeper wavelength dependence ($\sim \lambda^{-2}$) than amorphous carbon ($\sim \lambda^{-1.2}$). Because of the flatter wavelength dependence, amorphous carbon has a higher opacity at 10μm than graphite, which makes the silicate feature less pronounced. In order to improve the strength of this feature (to be consistent with SED data), we had to introduce significant flaring in the disk as this would expose more and more disk material to stellar radiation, which produces the pronounced silicate feature. However, such flared-disk geometry fails to reproduce the shape of the observed visibility spectrum. The best-fit model gives unacceptably large chi-square value ($\chi^2_{IF} \sim 56$) for our interferometric data although the fit to SED data is reasonable ($\chi^2_{SED} \sim 2$). Hence we ruled-out silicate-amorphous carbon grain composition for the disk of MWC 325. A silicate only model simultaneously fits the visibility and SED data. However, such as grain composition is far from that of the interstellar grains. Hence we did not consider this case in our work.

The decomposition of the N-band silicate feature of MWC 325 shows a strong dominance of 2 μm amorphous silicate grains with a mass fraction of 0.85 (Schütz et al. 2009). Sub-micron (0.1 μm) amorphous silicate and enstatite and 2 μm silica grains are present in much smaller amounts. More recently, Juhasz et al. (2010) decomposed mid-infrared spectra (5.5—37 μm) of a sample of Herbig Ae/Be stars taken with the *Spitzer Space Telescope*. They used a mixture of five-grain species, namely, amorphous silicate with olivine and pyroxene stoichiometry, crystalline forsterite, and enstatite and silica, together with polycyclic aromatic hydrocarbons (PAHs) and report a mass fraction of 0.68 for micron-sized amorphous silicate grains. In our best model, the mass fraction of the 2μm grains is 0.94. Thus our model is qualitatively consistent with Juhasz et al. 2010 and Schütz et al. (2009), who both conclude that micron-sized amorphous silicate grains dominate in MWC 325. Direct comparison of our grain composition with these mid-infrared modeling papers is not possible since the mid-infrared decomposition method is sensitive only to dust grains from the surface of the disk that show resonances in the mid-infrared (slicate based grains). Thus featureless grains such as amorphous carbon or iron are unaccounted for here. High spectral resolution interferometric measurements in the mid-infrared are essential to study the presence of crystalline grains since they produce narrow features.

Earlier studies (Monnier & Millan-Gabet 2002; Eisner et al. 2004) using broad band, usually single-wavelength, interferometric data, recognized a difference in the near-infrared size vs. luminosity behavior of high luminosity objects (pre-main sequence Be) compared to lower luminosity ones (pre-main sequence Ae), the former being more consistent with "classical disk" models. This has been revisited most recently by Vinković & Jurkić (2007), who use a model-independent comparison of visibility to scaled baseline and find a distinction between low-luminosity ($\leq 10^3$ $L_\odot$) and high-luminosity ($\geq 10^3$ $L_\odot$) YSO disks where the luminosity break point corresponds to an approximate spectral type B3-B5. These authors modeled the visibility clusters of low luminosity Herbig Ae/Be stars with optically thick rings of 0°–60° inclination at a dust sublimation temperature of ~1500K. The alternate model used to fit these visibility clusters was a dusty halo model with optical depths of ~0.15–0.8. They modeled visibility clusters of high luminosity Herbig Be stars with a classical accretion disk model and T Tauris stars with a dusty halo model.

However, multi-wavelength interferometric studies have not always supported these conclusions when objects are modeled in detail (Kraus et al. 2008a; Acke et al. 2008; Ragland et al. 2009). Our earlier multi-wavelength results (Ragland et al. 2009) show that MWC 419 (B8, 330 $L_\odot$) has disk characteristics of a high-luminosity object in the categories of Monnier & Millan-Gabet (2002).

However, it has a luminosity below the break point of $10^3$ $L_\odot$ identified by Vinković & Jurkić (2007) and fits within the population of lower luminosity Herbig starts in their model-independent comparison. Their physical interpretation of the low-luminosity Herbig group is an optically thick disk with an optically thin dust sublimation cavity and an optically thin dusty outflow.

MWC 325 (A2, 83 $L_\odot$) falls into the low luminosity group in the classification of Vinković & Jurkić (2007). Our results presented in this paper do not support their conclusion of a ring or halo model for this low luminosity YSO disk. While, a ring model could reproduce, to first order, the bright inner ring region seen from the *K*-band measurements, it would have difficulties in explaining the extended disk structures seen from the *L*-band and *N*-band measurements. The apparent discrepancies between our results and that of Vinković & Jurkić (2007) could be attributed to the fact that their conclusions are based on near-infrared interferometric data while this work incorporated a broader spectral region.

MWC 325 is in the Group II category of Herbig AeBe stars (Acke et al. 2009), in the classification scheme of Meeus et al. (2001), with a relatively weak IR excess and PAH emission compared to the Group I category. These studies based on Spitzer measurements suggest that the disk of MWC 325 is a self-shadowed disk - with an inner rim that is blocking star light from reaching the outer regions of the disk. Our finding of disk geometry with little or no flaring is consistent their results.

## 5. SUMMARY

This article reports the first milliarcsecond angular resolution *N*-band nulling and *L*-band $V^2$ observations of a Herbig Ae star (MWC 325), along with *K*-band $V^2$ data, all spectrally dispersed. This multi-wavelength observational capability is well suited to probing the temperature distribution in the inner regions of YSO disks, which is very important for distinguishing different models and gaining insight into the three dimensional geometry of the inner disk (see also Ragland et al. 2009).

A simple pole-on uniform disk model was used to infer an increase in size from 2–12 μm, confirming that the disk is extended with a radial temperature gradient. Notably, there is no difference in the wavelength trend of the visbilities within the broad 10 micron silicate feature compared to the adjacent continuum. We find that the classical accretion disk and the power-law temperature gradient models fail to fit simultaneously both interferometric and SED data. A 2-D slightly shadowed-disk radiative transfer model fits the spectrally dispersed interferometric measurements and the SED data reasonably well. This model implies that the disk surrounding MWC 325 is a nearly flat-disk with if at most only slight flaring in the outer regions of the disk in contrast to other intermediate mass Ae disks such as MWC 275 and AB Aur, and to their more massive counterparts, such as MWC 419, which exhibit a flat geometry. The dominance of sub-micron grains and the absence of significant flaring in the disk of MWC 325 found from our study suggest that dust grain growth and dust sedimentation has occurred in the disk of MWC 325. A more complete sample of YSO disk observations with adequate wavelength and (u,v) coverage, plus detailed radiative transfer modeling, are required to address the intriguing inner disk geometry of these sources and to address the structural differences between the Herbig Ae and Be disks.

## ACKNOWLEDGEMENTS

The Keck Interferometer is funded by the National Aeronautics and Space Administration (NASA). Observations presented were obtained at the W. M. Keck Observatory, which is operated as a scientific partnership among the California Institute of Technology, the University of California, and NASA. The Observatory was made possible by the generous financial support of the W. M. Keck Foundation. KI observations were taken through Keck Director's time. We thank E. Appleby, A. Cooper, C. Felizardo, J. Herstein, D. Medeiros, D. Morrison, T. Panteleeva, B. Parvin, B. Smith, K. Summers, K. Tsubota, C. Tyau, E. Wetherell, P. Wizinowich, and J. Woillez for their contributions to KI operations. CHARA observations were taken through NOAO TAC time. We thank Gail Schaefer, P.J. Goldfinger, Chris Farrington and Theo ten Brummelaar for support of the CHARA observing program and Tabetha Boyajian for advice on data reductions. STR acknowledges partial support from NASA grant NNH09AK731.